\documentclass[prb,superscriptaddress,amsmath,amssymb,floatfix,showpacs,twocolumn]{revtex4-1}

\usepackage{amsfonts,amsmath,amssymb}
\usepackage{epsfig}
\usepackage{times}
\usepackage{graphicx}
\usepackage{bm}
\usepackage{xcolor}
\usepackage{dcolumn}
\usepackage{array}
\usepackage{float}
\usepackage{soul}

\usepackage{feynmp-auto}
\usepackage{cases}

\newcommand\tmax{t_{\text{max}}}

\begin{document}

\title{Cancellation of vacuum diagrams and long-time limit in out-of-equilibrium diagrammatic Quantum Monte Carlo} 

\author{Alice Moutenet}
\affiliation{CPHT, CNRS, Ecole polytechnique, IP Paris, F-91128 Palaiseau, France}
\affiliation{Coll\`ege de France, 11 place Marcelin Berthelot, 75005 Paris, France}
\affiliation{Center for Computational Quantum Physics, Flatiron Institute, 162 Fifth Avenue, New York, NY 10010, USA}
\author{Priyanka Seth}
\affiliation{Institut de Physique Th\'eorique (IPhT), CEA, CNRS, UMR 3681, 91191 Gif-sur-Yvette, France}
\author{Michel Ferrero}
\affiliation{CPHT, CNRS, Ecole polytechnique, IP Paris, F-91128 Palaiseau, France}
\affiliation{Coll\`ege de France, 11 place Marcelin Berthelot, 75005 Paris, France}
\author{Olivier Parcollet} 
\affiliation{Center for Computational Quantum Physics, Flatiron Institute, 162 Fifth Avenue, New York, NY 10010, USA}
\affiliation{Institut de Physique Th\'eorique (IPhT), CEA, CNRS, UMR 3681, 91191 Gif-sur-Yvette, France}

\date{\today}

\begin{abstract}
  We express the recently introduced  real-time diagrammatic Quantum Monte Carlo, 
{\it Phys. Rev. B \textbf{91}, 245154 (2015)},
in the Larkin-Ovchinnikov basis in Keldysh space.
Based on a perturbation expansion in the local interaction $U$, the special form
 of the interaction vertex allows to write diagrammatic rules in which vacuum Feynman diagrams
directly vanish. This reproduces the main property of the previous algorithm, without the cost of
the exponential sum over Keldysh indices.
  In an importance sampling procedure, this implies that only interaction times in the vicinity of the measurement time contribute.
  Such an algorithm can then directly address the long-time limit needed in the study of steady states in out-of-equilibrium systems.
We then implement and discuss different variants of Monte Carlo algorithms in the Larkin-Ovchinnikov basis.
A sign problem reappears, showing that the cancellation of vacuum diagrams has no direct impact on it.
\end{abstract}

\maketitle

\newpage


\section{Introduction}

The development of high-precision and controlled computational methods for non-equilibrium
models in strongly-correlated regimes is  a subject of growing
interest in theoretical condensed-matter physics.
Recent years have seen significant experimental progress
with quantum transport through mesoscopic systems \cite{Potok_2007},  metal-insulator transitions driven by
an electric field \cite{Nakamura2013} or light-induced
superconductivity \cite{Fausti2011, Nicoletti2014, Casandruc2015, Nicoletti2016, Nicoletti2018}.

Powerful tools have been designed for the study of quantum systems at
equilibrium. Notably, the combination of dynamical mean-field theory \cite{RevModPhys.68.13, Kotliar2006, Aoki2014} and
state-of-the-art continuous-time Quantum Monte Carlo (QMC) algorithms such as
CT-INT \cite{Rubtsov2004, rubtsov_prb_2005}, CT-AUX \cite{Gull2008}, or CT-HYB
\cite{Werner_PRL_2006,Werner_PRB_2007} have allowed for great advances. When
considering out-of-equilibrium systems, however, early attempts to construct
similar perturbation-expansion-based real-time QMC algorithms encountered  an
exponential sign problem that prevented them from reaching long times and large
interactions \cite{Muhlbacher2008, Werner2009, Werner2010, Schiro2009,
Schiro2010}. Other approaches such as the density matrix renormalization group
(DMRG) \cite{White1992, White1993, Schollwoeck2005} also struggle in the long-time limit due to entanglement growth. There is therefore still a great
need for high-precision numerical methods that would be able to access the
non-equilibrium steady states of strongly-interacting quantum systems.


Current efforts to build real-time quantum Monte Carlo methods mainly explore two routes: the inchworm  algorithm \cite{Cohen2014a, Cohen2014b, Cohen2015, Chen2017a, Chen2017b, Antipov2017, Boag2018} and the so-called ``diagrammatic''
QMC \cite{Profumo_prb_2015, Bertrand2019a, Bertrand2019b} which is the subject of this article. Using an expansion of physical quantities in powers of the interaction $U$, this algorithm
has been shown to directly address the infinite-time steady states.
The name ``diagrammatic'' refers to its imaginary-time counterparts that were historically constructing a Markov chain in the space of
Feynman diagrams \cite{Prokofev1998, Prokofev2007, vanhoucke_2010, Bourovski2004}.

First introduced in Ref. \onlinecite{Profumo_prb_2015}, the real-time diagrammatic QMC algorithm stochastically samples
physical quantities using an importance sampling. At a given perturbation order $n$, its key idea is to regroup a factorial
number of Feynman diagrams in a sum over Keldysh indices of $2^n$ determinants.
This exponential sum has been shown to cancel vacuum diagrams, a property also used in recent diagrammatic QMC methods
in imaginary-time \cite{riccardo_2017, Alice_2018, fedor_2017}. As a direct consequence, the Monte Carlo sampling only involves interaction times in a neighborhood around
the measurement time $\tmax$: we talk about the \emph{clusterization} of times.
The computation of the Monte Carlo weight is exponential in the perturbation order
but {\it uniform in time}, at any temperature.
The algorithm can therefore address long, even infinite, times in the computation of contributions to the perturbation theory.
This method was recently generalized to compute the Green's function
and tested in quantum impurity models \cite{Bertrand2019a, Bertrand2019b}. 
The current form of the algorithm is able to 
compute the Kondo resonance at low temperature in the strongly-correlated Kondo regime.

Coefficients of the  expansion being written in terms of high-dimensional integrals of the sum of determinants,
its exponential scaling limits our capability to compute high orders
with great precision (we typically are limited to 10 of them). Even though non-perturbative information and Bayesian techniques can overcome noise amplification
occurring in the resummation of the series~\cite{Bertrand2019b}, this can prevent the algorithm to reach very large $U$.

In this article, we show that we can obtain the cancellation
of diagrams and the long-time clusterization property {\it without} summing
an exponential number of terms.  Using the Larkin-Ovchinnikov (LO) basis in Keldysh
space, we rewrite the integrand as a sum of $4^n$ determinants, but we show that
diagrammatic rules in this basis are such that \emph{every} diagram
has the clusterization property. In other words, the elimination of vacuum diagrams
is directly achieved in the diagrammatics without the need of an exponential sum.
We then implement and compare two Monte-Carlo algorithms based on this mathematical property. 
Both sample single determinants at a polynomial cost, but then one measures in the LO basis (LO algorithm) while the other
measures in the original basis (mixed algorithm).
We obtain that a simple implementation of the real-time diagrammatic QMC in the Larkin-Ovchinnikov
basis leads to a severe sign problem, which is reduced in the mixed algorithm.
This shows that the main effect of the exponential sum of determinants, beyond the cancellation of vacuum disconnected diagrams,
is to reduce the sign problem of this class of algorithms.

This article is organized as follows. First, we present in Sec. \ref{sec:keldysh_formalism} the usual Keldysh formalism in the $\{\pm\}$ basis,
briefly summarize the diagrammatic rules and then derive the cancellation of vacuum diagrams and the clusterization of the density when 
summing over Keldysh indices. We follow the same structure in Sec. \ref{sec:lo_formalism} where we introduce the Larkin-Ovchinnikov basis,
showing that all vacuum diagrams are equal to zero, so that density contributions directly clusterize around the measurement time.
We then detail in Sec. \ref{sec:implementation} the Monte Carlo implementation of the original algorithm presented in Ref. \onlinecite{Profumo_prb_2015}
($\pm$ algorithm) and two algorithms based on the Larkin-Ovchinnikov formalism (LO and mixed algorithms).
In Sec. \ref{sec:results} we compute the density of an impurity level coupled to a bath, present the results of all three algorithms and explain
the origin of the observed error bars. We finally conclude in Sec. \ref{sec:conclusion}.

%
%
\section{Keldysh formalism} \label{sec:keldysh_formalism}


We work in the Keldysh formalism \cite{Schwinger_1961, Keldysh, Rammer_rmp_1986, 2009_keldysh_review}. In this framework, operators act on the Keldysh contour $\mathcal{C}$
consisting of a forward branch, from an initial time $t_0$ (that we take equal to $0$ in the following) to a given time $\tmax$,
and a backward branch, from $\tmax$ to $t_0$.
The system is initially prepared at equilibrium without interactions.
A Keldysh point $k$ on $\mathcal{C}$ is defined as  a
pair $k \equiv (t, \alpha)$ with a time $t \in [0, \tmax]$ and a Keldysh index $\alpha \in \{\pm\}$ indicating which branch is to be considered.
The + (resp. -) index denotes the forward (resp. backward) branch, as depicted below.

\begin{center}
\begin{fmffile}{keldysh_contour}
\begin{fmfgraph*}(200,10)
\fmfset{arrow_len}{3mm}
\fmfleft{i1,i2,i3}
\fmfright{o1,o2,o3}
\fmf{phantom}{i2,v1}
\fmf{phantom,label.dist=0,tension=15.}{v1,o2}
\fmf{phantom}{v1,v2}
\fmf{phantom,label.dist=0,label= \vspace{-.8cm} \hspace{-.5cm}$\tmax$}{v2,o2}
\fmf{plain_arrow, label.dist=0, label= \hspace{-1.2cm} \vspace{-.4cm}$\mathrm{\textbf{-}}$}{o1,i1}
\fmf{plain_arrow, label.dist=0, label= \hspace{1.2cm} \vspace{.4cm}$\mathrm{\textbf{+}}$}{i3,o3}
\fmf{phantom,tag=1, label.dist=0, label= \vspace{-.75cm} \hspace{-.5cm} $0$}{i1,i3}
\fmfipath{p[]}
\fmfiset{p1}{vpath1(__i1,__i3)}
\fmfi{plain}{subpath (length(p1)/3, 2length(p1)/3) of p1}
\fmffreeze
\fmf{plain}{i2,v2}
\fmf{phantom_arrow}{v1,o2}
\fmf{plain,right}{o1,o3}
\end{fmfgraph*}
\end{fmffile}
\end{center}

Note that both branches are along the real axis and are displaced only for graphical purposes.
In the following, Greek letters refer to $\pm$ indices unless otherwise stated.
We define a contour operator $T_{\mathcal{C}}$ that follows the arrows on the above picture: $T_\mathcal{C}$ coincides
with the usual time-ordering operator T on the $+$ branch, with the anti-time ordered operator \v{T} on the $-$ branch, and considers all Keldysh points on the backward branch to be later than points on the forward branch.

The formalism we develop in this section is valid for any general model described by a noninteracting Green's function $g$ and a density-density interaction.
However, for the sake of simplicity, we consider interacting electrons on a single energy level. The operator $c_\sigma$ (resp. $c_\sigma^\dagger$) destroys
(resp. creates) an electron with spin $\sigma = \uparrow, \downarrow$. The interaction term, turned on at $t=0$, is given by
the interaction vertex $U n_\uparrow n_\downarrow$, where 
$n_\sigma \equiv c_\sigma^\dagger c_\sigma$ is the density operator.

We define the time-dependent Green's function
\begin{equation} \label{gf_def}
\hat G_\sigma(t,t') \equiv -i \langle T_{\mathcal{C}}  c_\sigma(t) c_\sigma^\dagger(t')\rangle,
\end{equation}
where $c_\sigma^{(\dagger)}(t)$ is the Heisenberg representation of $c_\sigma^{(\dagger)}$ and the average
is taken with respect to the initial noninteracting state.
The Green's function takes the form of a $2 \times 2$ matrix in the $\{\pm\}$ basis: $\hat G_\sigma = \begin{pmatrix}
G_\sigma^{++} & G_\sigma^< \\
G_\sigma^> & G_\sigma^{--}
\end{pmatrix}$,
where
\begin{subequations}
\begin{align}
  G_\sigma^<(t,t')  &\equiv i\langle c_\sigma^\dagger(t') c_\sigma(t) \rangle , \\
  G_\sigma^>(t,t')  &\equiv -i \langle c_\sigma(t) c_\sigma^\dagger(t')\rangle , \\
  G_\sigma^{++}(t,t') &\equiv -i \langle \text{T} c_\sigma(t) c_\sigma^\dagger(t') \rangle , \\
  G_\sigma^{--}(t,t')  &\equiv -i \langle \text{\v{T}} c_\sigma(t) c_\sigma^\dagger(t') \rangle .
\end{align}
\end{subequations}
Throughout the article, noninteracting Green's functions will be denoted by lower case letters, interacting ones by upper case letters,
and a $~\hat{}~$ denotes a matrix.

\subsection{Diagrammatic rules}

In this article, we construct perturbation series in the interaction $U$ for
physical observables of interest.  Computing contributions at
different perturbation orders relies on the evaluation of Feynman diagrams
obeying rules that we briefly summarize.

A straight line represents a noninteracting Green's function
\begin{fmffile}{keldysh_g0}
  \begin{equation}\label{keldysh_g0}
\begin{gathered}
\begin{fmfgraph*}(40,20)
\fmfset{arrow_len}{3mm}
\fmfleft{v1}
\fmfright{v2}
\fmf{plain_arrow, label.dist=0, label= \vspace{.5cm} \hspace{-.4cm}$\sigma$}{v1,v2}
\fmflabel{$t', \beta$}{v1}
\fmflabel{$t, \alpha$}{v2}
\end{fmfgraph*}
\end{gathered}
\hspace{.9cm}
= i \left(\hat g_{\sigma}\right)_{\alpha\beta} (t,t').
\end{equation}
\end{fmffile}
\hspace{-.15cm}Because the interaction has the form $Un_\uparrow n_\downarrow$, an interaction vertex is characterized by a single Keldysh point
$\{t, \alpha\}$, and the
indices of the four legs all have to be equal to the Keldysh index $\alpha$ \\
\begin{fmffile}{keldysh_vertex}
  \begin{equation}\label{keldysh_vertex}
\begin{gathered}
\begin{fmfgraph*}(40,30)
\fmfset{arrow_len}{3mm}
\fmfleft{i1,i2}
\fmfright{o1,o2}
\fmf{plain_arrow}{i1,v}
\fmf{plain_arrow}{v,o2}
\fmf{plain_arrow}{o1,v}
\fmf{plain_arrow}{v,i2}
\fmfdot{v}
\fmflabel{$\alpha$, $\uparrow$}{i1}
\fmflabel{$\alpha$, $\downarrow$}{o1}
\fmflabel{$\alpha$, $\uparrow$}{i2}
\fmflabel{$\alpha$, $\downarrow$}{o2}
  \fmflabel{$\{t, \alpha\}$}{v}
\end{fmfgraph*}
\end{gathered}
\hspace{0.5cm}
=-i\alpha U.
\end{equation}
\end{fmffile} \\
\hspace{-.09cm}Hence, for every interaction time $t$, there are two possible vertices.
The sum of the different $\{\pm\}$ configurations can be written in the $\mathcal{H}_\uparrow \otimes \mathcal{H}_\downarrow$ space,
in the form
\begin{equation} \label{4_leg_pm}
-iU \left( m_+ \otimes m_+ - m_- \otimes m_-\right),
\end{equation}
where $m_+ = \begin{pmatrix}
1 & 0 \\
0 & 0
\end{pmatrix}$ and $m_- = \begin{pmatrix}
0 & 0 \\
0 & 1
\end{pmatrix}$ are matrices in the $\{\pm\}$ basis, and $\mathcal{H}_\sigma$ is the Hilbert space for spin $\sigma$.
Furthermore, an interaction of the form $hc_\sigma^\dagger c_\sigma$ in the Hamiltonian would give rise
to 2-leg vertices of the form
\begin{fmffile}{keldysh_2_leg}
\begin{equation}
\begin{gathered}
\begin{fmfgraph*}(40,20)
\fmfset{arrow_len}{3mm}
\fmfleft{i1}
\fmfright{o1}
\fmf{plain_arrow,tension=2,label.dist=0, label=\vspace{.3cm} \hspace{-.8cm} $\sigma$}{i1,v1}
  \fmf{plain_arrow,tension=2,label.dist=0, label=\vspace{-.6cm} \hspace{-.9cm} $\{t,, \alpha\}$}{v1,o1}
\fmffreeze
\fmf{phantom,label.dist=0, label=\vspace{.6cm} \hspace{-.9cm} $h$}{v1,o1}
\fmflabel{$\alpha$}{i1}
\fmflabel{$\alpha$}{o1}
\fmfv{decor.shape=pentagram,decor.filled=full, decor.size=2thick}{v1}
\end{fmfgraph*}
\end{gathered}
\hspace{0.5cm}
= -i\alpha h.
\end{equation}
\end{fmffile}
\hspace{-0.09cm}These do not appear directly in the diagrammatics but will be formally useful when deriving the expression of the fermionic bubble.
The sum over Keldysh indices reads
  $-ih (m_+ - m_-)$
in both $\mathcal{H}_\uparrow$ and $\mathcal{H}_\downarrow$ spaces.

With the expression of the 4-leg interaction vertex, the following fermionic bubble reads
\begin{fmffile}{keldysh_closed_loop}
\begin{equation} 
\begin{gathered}
\begin{fmfgraph*}(50,20)
\fmfset{arrow_len}{3mm}
\fmfleft{i1,i2}
\fmfright{o1,o2}
\fmf{plain_arrow,tension=2,label.dist=0, label=\vspace{.3cm} \hspace{-.8cm} $\sigma$}{i1,v1}
\fmf{plain_arrow,tension=2}{v1,o1}
\fmf{phantom}{i2,v2,o2}
  \fmflabel{$\{t, \alpha\}$}{v1}
\fmflabel{$\alpha$}{i1}
\fmflabel{$\alpha$}{o1}
\fmfdot{v1}
\fmffreeze
\fmf{plain_arrow,right,tension=2}{v1,v2,v1}
\fmf{phantom,label.dist=0, label=$\bar{\sigma}$}{v1,v2}
\end{fmfgraph*}
\end{gathered}
\hspace{0.5cm}
= \alpha Ug^{\alpha\alpha}_{\bar\sigma}(t,t).
\end{equation}
\end{fmffile}
\hspace{-0.09cm}
Because of the form of the interaction term, we have 
\begin{equation}\label{eq_time_pm}
g_\sigma^{\alpha\alpha}(t,t) = g_\sigma^<(t,t).
\end{equation}
Hence, the above diagram reduces to $\alpha U g_{\bar{\sigma}}^<(t,t)$,
which can be formulated as a 2-leg vertex with a $iU g_{\bar\sigma}^<(t,t)$ field
\begin{fmffile}{keldysh_closed_loop2}
\begin{equation}\label{closed_vertex_eq}
\begin{gathered}
\begin{fmfgraph*}(50,20)
\fmfset{arrow_len}{3mm}
\fmfleft{i1,i2}
\fmfright{o1,o2}
\fmf{plain_arrow,tension=2,label.dist=0, label=\vspace{.3cm} \hspace{-.8cm} $\sigma$}{i1,v1}
\fmf{plain_arrow,tension=2}{v1,o1}
\fmf{phantom}{i2,v2,o2}
  \fmflabel{$\{t, \alpha\}$}{v1}
\fmflabel{$\alpha$}{i1}
\fmflabel{$\alpha$}{o1}
\fmfdot{v1}
\fmffreeze
\fmf{plain_arrow,right,tension=2}{v1,v2,v1}
\fmf{phantom,label.dist=0, label=$\bar{\sigma}$}{v1,v2}
\end{fmfgraph*}
\end{gathered}
\hspace{0.5cm}
= 
\hspace{0.7cm}
\begin{gathered}
\begin{fmfgraph*}(40,20)
\fmfset{arrow_len}{3mm}
\fmfleft{i1}
\fmfright{o1}
\fmf{plain_arrow,tension=2,label.dist=0, label=\vspace{-.4cm} \hspace{-.8cm} $\sigma$}{i1,v1}
  \fmf{plain_arrow,tension=2,label.dist=0, label=\vspace{-.6cm} \hspace{-.9cm} $\{t,,\alpha\}$}{v1,o1}
\fmffreeze
\fmf{phantom,label.dist=0, label=\vspace{.6cm} \hspace{-.9cm} $iUg_{\bar\sigma}^<(t,,t)$}{v1,o1}
\fmflabel{$\alpha$}{i1}
\fmflabel{$\alpha$}{o1}
\fmfv{decor.shape=pentagram,decor.filled=full, decor.size=2thick}{v1}
\end{fmfgraph*}
\end{gathered}
\end{equation}
\end{fmffile}

If $\mathcal{M}$ is the quantity we want to compute (later on the density), its perturbation expansion
is given by $\mathcal{M} = \sum_n \mathcal{M}_n U^n$. Because of the form of the interaction
vertex, we have
\begin{align} \label{m_in_U}
  \mathcal{M}_n &= \int_\mathcal{C} \mathrm{d}k_1 \dots \mathrm{d}k_n \text{ } \mathcal{M}_n^\pm(k_1, \dots, k_n) \\
  &= \int_0^{\tmax} \mathrm{d} t_1 \dots \mathrm{d}t_n \sum_{\alpha_1 \dots \alpha_n} \mathcal{M}_n^\pm(\{t_i,\alpha_i\}_{1 \leq i \leq n}),
\end{align}
where $ \mathcal{M}_n^\pm(\{t_i,\alpha_i\}_{1 \leq i \leq n})$ can be expressed as a product of determinants, their precise form depending on the
measured quantity. Throughout this paper, the $\pm$ superscript will denote quantities expressed in the $\{\pm\}$ basis.
Moreover times integrated over are always considered ordered.

\subsection{Cancellation of vacuum diagrams when summing over Keldysh indices}

Due to the forward-backward nature of the contour $\mathcal{C}$, the partition function $Z$ is exactly equal to 1 in the real-time Keldysh formalism.
Expressing $Z$ as a series in $U$ ($Z = \sum_n Z_n U^n$), this property implies that all $Z_n$
are vanishing for $n \geq 1$. Because of the form of Eq. \eqref{m_in_U}, this cancellation involves both the integral over times and the sum over Keldysh indices. However, it was proven
by Profumo and co-workers in Ref.  \onlinecite{Profumo_prb_2015} that only the latter is needed. For all $n \geq 1$, $\{t_1, \dots, t_n\} \in [0,\tmax]^n$,
\begin{equation}\label{CancellationZ_pm}
\sum_{\alpha_1 \dots \alpha_n} Z_n^\pm(\{t_i,\alpha_i\}_{1 \leq i \leq n}) = 0,
\end{equation}
where
\begin{equation}
  \begin{split}
    Z_n^\pm(\{t_i,\alpha_i\}_{1 \leq i \leq n})& = (-i\alpha_1) \dots (-i\alpha_n) \times i^n i^n \\
& \hspace{-.5cm} \times \prod_\sigma \det \left[\left(\hat g_\sigma\right)_{\alpha_i \alpha_j} (t_i, t_j)\right]_{1 \leq i,j \leq n}.
\end{split}
\end{equation}
Each $(-i\alpha_k)$ comes from Eq. \eqref{keldysh_vertex}, and the two $i^n$ factors from the fact that a straight line actually represents an $i\hat g$ (Eq. \eqref{keldysh_g0}).

For every configuration of times $\{t_1, \dots, t_n\}$, vacuum diagrams therefore cancel when performing
the explicit $2^n$ sum over Keldysh indices. 
Recent developments in imaginary-time diagrammatic QMC also achieved, through an iterative procedure, the cancellation of vacuum (and, later on, non one-particle irreducible) diagrams 
at every Monte Carlo step at an exponential cost in the perturbation order. \cite{riccardo_2017, Alice_2018,fedor_2017}

\subsection{Density computation and clusterization}

In the following, we compute the density $d$ of electrons with spin $\uparrow$
on the impurity level at the end point of the Keldysh contour, $d \equiv \langle n_\uparrow(\tmax)\rangle$.

In the $\{\pm\}$ basis, let us note that $d = (\hat G_\uparrow)_{01}(\tmax, \tmax) / i$.
Hence we can represent the measurement vertex as a ``special" vertex bearing time $\tmax$, such
that the ingoing and outgoing Keldysh indices are 0 and 1:
\begin{fmffile}{keldysh_measurement}
\begin{equation}
\begin{gathered}
\begin{fmfgraph*}(50,20)
\fmfset{arrow_len}{3mm}
\fmfleft{i1}
\fmfright{o1}
\fmf{dashes_arrow,tension=2,label.dist=0, label=\vspace{.4cm} \hspace{.1cm} $0$}{i1,v1}
\fmf{dashes_arrow,tension=2,label.dist=0, label=\vspace{-.7cm} \hspace{-.9cm} $\tmax$}{v1,o1}
\fmffreeze
\fmf{phantom,label.dist=0, label=\vspace{.45cm} \hspace{-.5cm} $1$}{v1,o1}
\fmfv{decor.shape=circle,decor.filled=shaded, decor.size=4thick}{v1}
\end{fmfgraph*}
\end{gathered}
\end{equation}
\end{fmffile}
\hspace{-.09cm}Note that the surrounding lines are dashed because they should bear a $\hat{g}$ propagator (instead of an $i\hat g$ one as in the rest of the formalism).
The order-$n$ contribution to $d$ reads
\begin{equation} \label{d_keldysh}
\begin{split}
d_n = & \int_0^{\tmax} \mathrm{d}t_1 \dots \mathrm{d} t_n \sum_{\alpha_1 \dots \alpha_n}(-i\alpha_1)\dots(-i\alpha_n) \\
& \times \frac{i^{n+1}i^n}{i^2}\prod_\sigma \det \mathcal{D}_\sigma^\pm(\{t_i, \alpha_i\}_{1 \leq i \leq n}),
\end{split}
\end{equation}
where 
\begin{widetext}
\begin{equation}\label{D_up_pm}
\mathcal{D}_\uparrow^\pm(\{t_i, \alpha_i\}_{1 \leq i \leq n}) = 
\begin{pmatrix}
 \left[(\hat g_\uparrow)_{\alpha_i \alpha_j}(t_i,t_j)\right]_{1 \leq i,j \leq n} & \begin{matrix}
(\hat g_\uparrow)_{\alpha_1 1}(t_1, \tmax) \\
\vdots \\
(\hat g_\uparrow)_{\alpha_n 1}(t_n, \tmax)
\end{matrix} \\
\begin{matrix}
(\hat g_\uparrow)_{0\alpha_1}(\tmax, t_1) & \dots & (\hat g_\uparrow)_{0\alpha_n}(\tmax, t_n)
\end{matrix} & (\hat g_\uparrow)_{01}(\tmax, \tmax)
\end{pmatrix},
\end{equation}
\end{widetext}
and
\begin{equation}
\mathcal{D}_\downarrow^\pm(\{t_i, \alpha_i\}_{1 \leq i \leq n}) = \left[ (\hat g_\downarrow)_{\alpha_i \alpha_j}(t_i,t_j)\right]_{1 \leq i,j \leq n}.
\end{equation}

Using the cancellation of vacuum diagrams when summing over Keldysh indices, we reproduce in Appendix \ref{app:clusterization} the argument of Ref. \onlinecite{Profumo_prb_2015}
showing that the computation of $d_n$ only involves the sampling of interaction times close to $\tmax$.
As a direct consequence, Monte Carlo algorithms implementing this sum in the calculation of the weight can address any 
measurement time $\tmax$,
when earlier methods were limited to short-term measurements \cite{Muhlbacher2008, Werner2009, Werner2010, Schiro2009, Schiro2010}.
We talk about the \emph{clusterization} of interaction times in the computation of the density.

\section{Larkin-Ovchinnikov Formalism}\label{sec:lo_formalism}

Starting from the expression of the Green's function in the $\{ \pm \}$ basis, we define its counterpart in the LO basis, $\hat G^{\text{LO}}$, through the following transformation  \cite{Keldysh, LO}
\begin{equation}
\hat G_\sigma^{\text{LO}}(t,t') \equiv L^\dagger \tau_3 \hat G_\sigma(t,t') L,
\end{equation}
where $L = \frac{1}{\sqrt{2}} \begin{pmatrix}
1 & 1 \\
-1 & 1
\end{pmatrix}$ and $\tau_3 = \begin{pmatrix}
1 & 0 \\
0 & -1
\end{pmatrix}$. The Green's function now takes the $2 \times 2$ form $\hat G_\sigma^{\text{LO}} = \begin{pmatrix}
R_\sigma & K_\sigma \\
0 & A_\sigma
\end{pmatrix}$,
where $R$, $K$, and $A$ are respectively the retarded, Keldysh and advanced Green's functions defined as
\begin{subequations}
\begin{align}
R_\sigma(t,t') &\equiv -i\theta(t-t') \langle \{c_\sigma(t), c_\sigma^\dagger(t')\} \rangle, \\
A_\sigma(t,t') &\equiv  i\theta(t'-t) \langle \{c_\sigma(t), c_\sigma^\dagger(t')\} \rangle, \\
K_\sigma(t,t') &\equiv -i \langle [c_\sigma(t), c_\sigma^\dagger(t')]\rangle.
\end{align}
\end{subequations}
In this basis, the Keldysh index $\alpha \in \{\pm\}$ is replaced by an LO index 0 or 1.
In the following, $l$ will always denote such an index unless otherwise stated.

\subsection{Diagrammatic rules}

To expose the diagrammatic rules in this formalism, let us first determine from Eq. \eqref{4_leg_pm} the form of the 4-leg interaction vertex in the LO basis.
The $m_+$ and $m_-$ matrices transform as
\begin{subequations}
\begin{gather}
  \label{mp_eq} L^\dagger \tau_3 m_+ L = \frac{1}{2} \begin{pmatrix}
1 &1 \\
1 & 1
\end{pmatrix} \equiv \frac{1}{2} \tau_\uparrow, \\
L^\dagger \tau_3 m_- L = \frac{1}{2} \begin{pmatrix}
-1 & 1 \\
1 & -1
\end{pmatrix} \equiv \frac{1}{2} \tau_\downarrow.
\end{gather}
\end{subequations}
Hence the sum of different LO contributions can be written
\begin{equation} \label{4_leg_lo}
-\frac{iU}{4} \left( \tau_\uparrow \otimes \tau_\uparrow - \tau_\downarrow \otimes \tau_\downarrow \right)
= -\frac{iU}{2} \left( \mathbf{1} \otimes \tau_\downarrow + \tau_\uparrow \otimes \mathbf{1} \right),
\end{equation}
where $\mathbf{1}$ is the $2 \times 2$ identity matrix.
Note that this is consistent with the symmetric form $-\frac{iU}{2}(\mathbf{1}\otimes\tau + \tau\otimes\mathbf{1})$ noted in Ref. \onlinecite{olivier_prb_2002},
where $\tau = \begin{pmatrix}
  0 & 1 \\
  1 & 0
\end{pmatrix}$.
The rhs form of Eq. \eqref{4_leg_lo} is the one we will retain in the rest of this article. We will show in Sec. \ref{subsec:cancellation} and \ref{subsec:density}
that the identity part of the vertex is essential in the proof of the cancellation of vacuum diagrams
and the clusterization of times in the computation of observables.

The key point of this expression of the vertex is that we can reduce the number of indices involved in the diagrammatics
using the fact that $\tau_\uparrow$ and $\tau_\downarrow$ are rank-1 matrices:
$\tau_\uparrow = v_\uparrow v_\uparrow^\top$ with $v_\uparrow = \begin{pmatrix} 1 \\ 1 \end{pmatrix} $ and $\tau_\downarrow = v_\downarrow (-v_\downarrow^\top)$ with $v_\downarrow = \begin{pmatrix} 1 \\ -1 \end{pmatrix} $.
We can therefore absorb the $\tau_\sigma$ part of the vertex in a redifinition of the noninteracting propagator (see below). 

An LO vertex can then be characterized by a tuple $\{ t, i_{\tau}, l\}$, where $t \in [0, \tmax]$,
$i_\tau \in \{-1,1\}$ and $l \in \{0,1\}$. $i_\tau = 1$ (resp. -1) indicates that the
$\uparrow$ (resp. $\downarrow$) spin is carrying the $\tau_\uparrow$ (resp. $\tau_\downarrow$) side, and $l$ is the LO index entering the identity-part of the vertex.
We store the information about both the bare propagator $\hat g_\sigma^{\mathrm{LO}} = \begin{pmatrix}
r_\sigma & k_\sigma \\
0 & a_\sigma
\end{pmatrix}$ and the nature of the vertices it is connected to in the form of a $3 \times 3$ matrix $\hat {\tilde{g}}_\sigma$. The two first indices corresponds to
a connection to the identity (with $l=$ 0 or 1), and the third one to the connection to a $\tau_\sigma$:
\begin{subequations}
\begin{align}
\left(\hat{\tilde{g}}_\sigma\right)_{ll'} &= \left(\hat g_\sigma^{\mathrm{LO}}\right)_{ll'}, \\
 \left(\hat{\tilde{g}}_\sigma\right)_{l2} &= \left(\hat g_\sigma^{\mathrm{LO}} v_\sigma \right)_l, \\
 \left(\hat{\tilde{g}}_\sigma\right)_{2l} &= \left(\sigma v_\sigma^\top \hat g_\sigma^{\mathrm{LO}}\right)_{l}, \\
\left(\hat{\tilde{g}}_\sigma\right)_{22} &=  \sigma v_\sigma^\top \hat g_\sigma^{\mathrm{LO}}v_\sigma,
\end{align}
\end{subequations}
with the convention that $\sigma = \uparrow$ should be understood as $+1$ and $\sigma = \downarrow$ as $-1$.

We obtain
\begin{equation}\label{tilde_g}
\hat{\tilde{g}}_\sigma = \begin{pmatrix}
r_\sigma & k_\sigma & r_\sigma +\sigma k_\sigma\\
0 & a_\sigma & \sigma a_\sigma \\
\sigma r_\sigma & \sigma k_\sigma + a_\sigma & \sigma[r_\sigma+ a_\sigma] + k_\sigma
\end{pmatrix}.
\end{equation}
To simplify upcoming equations, we express the indices of $\hat{\tilde g}_\uparrow$ and $\hat{\tilde g}_\downarrow$ at a vertex $\{t, i_\tau, l\}$ in the form of
two composite indices $L^\uparrow$ and $L^\downarrow$:
\begin{numcases}{L^\sigma =}
  2 & if $i_{\tau} = \sigma$ \nonumber \\
l & otherwise
\end{numcases}
Note that this $3 \times 3$ form of the Green's function comes from the absorption of the $\tau_\sigma$ part of the vertex
and has nothing to do with the  Baym-Kadanoff $L$-shaped contour used in thermal real-time computations.

With this notation, a straight line represents a noninteracting (modified)
Green's function
\begin{fmffile}{lo_g0}
  \begin{equation}\label{lo_g0}
\begin{gathered}
\begin{fmfgraph*}(40,20)
\fmfset{arrow_len}{3mm}
\fmfleft{v1}
\fmfright{v2}
\fmf{plain_arrow, label.dist=0, label= \vspace{.5cm} \hspace{-.4cm}$\sigma$}{v1,v2}
\fmflabel{$t', L'^\sigma$}{v1}
\fmflabel{$t, L^\sigma$}{v2}
\end{fmfgraph*}
\end{gathered}
\hspace{.9cm}
= i \left(\hat{\tilde g}_{\sigma}\right)_{L^\sigma L'^\sigma} (t,t').
\end{equation}
\end{fmffile}
\hspace{-.1cm}As discussed previously, the interaction vertex, proportional to the identity in the $\{ \pm\}$ basis, is now proportional to
$\frac{\mathbf{1} \otimes \tau_\downarrow + \tau_\uparrow \otimes \mathbf{1}}{2}$ in the $\mathcal{H}_\uparrow \otimes \mathcal{H}_\downarrow$ space
 \vspace{.4cm}
\begin{fmffile}{lo_vertex}
  \begin{equation}\label{lo_vertex}
\begin{split}
\begin{gathered}
\begin{fmfgraph*}(40,30)
\fmfset{arrow_len}{3mm}
\fmfleft{i1,i2}
\fmfright{o1,o2}
\fmf{plain_arrow}{i1,v}
\fmf{plain_arrow}{v,o2}
\fmf{plain_arrow}{o1,v}
\fmf{plain_arrow}{v,i2}
\fmfdot{v}
  \fmflabel{$\{t, i_\tau, l\}$}{v}
\fmflabel{$L^\uparrow, \uparrow$}{i1}
\fmflabel{$L^\downarrow, \downarrow$}{o1}
\fmflabel{$L^\uparrow, \uparrow$}{i2}
\fmflabel{$L^\downarrow, \downarrow$}{o2}
\end{fmfgraph*}
\end{gathered}
\hspace{0.8cm}
  = - \frac{i U}{2} & \left(\delta_{i_\tau 1}\delta_{L^\uparrow 2}\delta_{L^\downarrow l}\right. \\
  &  \left.  \hspace{.3cm}+ \delta_{i_\tau -1}\delta_{L^\downarrow 2}\delta_{L^\uparrow l}\right).
\end{split}
\end{equation}
\end{fmffile}
\hspace{-0.09cm}As $m_+ - m_-$ transforms into the $2 \times 2$ identity matrix in the LO basis, a 2-leg vertex is simply characterized by an interaction time $t$ and an LO index $l$.
A term $h c_\sigma^\dagger c_\sigma$ in the Hamiltonian would therefore give rise to the following vertex
\begin{fmffile}{lo_2_leg}
\begin{equation}
\begin{gathered}
\begin{fmfgraph*}(40,20)
\fmfset{arrow_len}{3mm}
\fmfleft{i1}
\fmfright{o1}
\fmf{plain_arrow,tension=2,label.dist=0, label=\vspace{.3cm} \hspace{-.8cm} $\sigma$}{i1,v1}
  \fmf{plain_arrow,tension=2,label.dist=0, label=\vspace{-.6cm} \hspace{-.9cm} $\{t,, l\}$}{v1,o1}
\fmffreeze
\fmf{phantom,label.dist=0, label=\vspace{.6cm} \hspace{-.9cm} $h$}{v1,o1}
\fmflabel{$L^\sigma$}{i1}
\fmflabel{$L^\sigma$}{o1}
\fmfv{decor.shape=pentagram,decor.filled=full, decor.size=2thick}{v1}
\end{fmfgraph*}
\end{gathered}
\hspace{0.7cm}
= -i h \delta_{L^\sigma l}.
\end{equation}
\end{fmffile}

With this expression of the interaction vertex, the  following fermionic bubble \hspace{.3cm}
\begin{fmffile}{lo_closed_loop}
\begin{fmfgraph*}(50,20)
\fmfset{arrow_len}{3mm}
\fmfleft{i1,i2}
\fmfright{o1,o2}
\fmf{plain_arrow,tension=2,label.dist=0, label=\vspace{.3cm} \hspace{-.8cm} $\sigma$}{i1,v1}
\fmf{plain_arrow,tension=2}{v1,o1}
\fmf{phantom}{i2,v2,o2}
  \fmflabel{$\{t, i_\tau,l\}$}{v1}
\fmflabel{$L^\sigma$}{i1}
\fmflabel{$L^\sigma$}{o1}
\fmfdot{v1}
\fmffreeze
\fmf{plain_arrow,right,tension=2}{v1,v2,v1}
\fmf{phantom,label.dist=0, label=$\bar{\sigma}$}{v1,v2}
\end{fmfgraph*}
\end{fmffile}
\hspace{.3cm}
evaluates to

\begin{equation}
  \begin{split}
    &   \frac{U}{2}  \delta_{L^\sigma 2}\left[ r_{\bar \sigma}(t,t) + a_{\bar \sigma}(t,t) \right]  \\
  +  &  \frac{U}{2} \delta_{L^\sigma l}\left[\bar \sigma r_{\bar \sigma}(t,t) + \bar \sigma a_{\bar \sigma}(t,t) + k_{\bar \sigma}(t,t)\right].
\end{split}
    \end{equation}
For the equal-time limit of the retarded, Keldysh and advanced Green's function, we choose 
a convention which ensures the consistency between the $\{\pm\}$ and LO basis.
We consider
\begin{subequations}\label{equal_time_choice}
\begin{gather}
k_\sigma (t,t) = 2 g_\sigma^{ <}(t,t), \\
r_\sigma (t,t) = a_\sigma (t,t) = 0,
\end{gather}
\end{subequations}
and we show that this is consistent with Eq. \eqref{eq_time_pm}.
Using Eq. \eqref{equal_time_choice}, the above fermionic bubble reduces to $Ug_{\bar \sigma}^< (t,t) \delta_{L^\sigma l}$.
It can be rewritten as a 2-leg vertex with a  $iU g_{\bar\sigma}^<(t,t)$ field
\begin{fmffile}{closed_vertex_lo}
\begin{equation}
\begin{gathered}
\begin{fmfgraph*}(50,20)
\fmfset{arrow_len}{3mm}
\fmfleft{i1,i2}
\fmfright{o1,o2}
\fmf{plain_arrow,tension=2,label.dist=0, label=\vspace{.3cm} \hspace{-.8cm} $\sigma$}{i1,v1}
\fmf{plain_arrow,tension=2}{v1,o1}
\fmf{phantom}{i2,v2,o2}
  \fmflabel{$\{t, i_\tau, l\}$}{v1}
\fmflabel{$L^\sigma$}{i1}
\fmflabel{$L^\sigma$}{o1}
\fmfdot{v1}
\fmffreeze
\fmf{plain_arrow,right,tension=2}{v1,v2,v1}
\fmf{phantom,label.dist=0, label=$\bar{\sigma}$}{v1,v2}
\end{fmfgraph*}
\end{gathered}
\hspace{0.5cm}
= 
\hspace{0.7cm}
\begin{gathered}
\begin{fmfgraph*}(40,20)
\fmfset{arrow_len}{3mm}
\fmfleft{i1}
\fmfright{o1}
\fmf{plain_arrow,tension=2,label.dist=0, label=\vspace{-.4cm} \hspace{-.8cm} $\sigma$}{i1,v1}
  \fmf{plain_arrow,tension=2,label.dist=0, label=\vspace{-.6cm} \hspace{-.9cm} $\{t,,l\}$}{v1,o1}
\fmffreeze
\fmf{phantom,label.dist=0, label=\vspace{.6cm} \hspace{-.9cm} $iUg_{\bar\sigma}^<(t,,t)$}{v1,o1}
\fmflabel{$L^\sigma$}{i1}
\fmflabel{$L^\sigma$}{o1}
\fmfv{decor.shape=pentagram,decor.filled=full, decor.size=2thick}{v1}
\end{fmfgraph*}
\end{gathered}
\end{equation}
\end{fmffile}
\\
\hspace{-.09cm}This equation is, up to a change of basis, the same as Eq. \eqref{closed_vertex_eq}. The choice of equal-time limit
described in Eq. \eqref{equal_time_choice} is therefore consistent with the $\{\pm\}$ basis formalism.

The order-$n$ contribution to the quantity $\mathcal{M}$ we want to measure in an expansion in $U$
is similar to Eq. \eqref{m_in_U}, but has to take account of the new form of the vertex
\begin{equation}
\mathcal{M}_n = \int_0^{\tmax} \mathrm{d}t_1 \dots \mathrm{d}t_n \sum_{\substack{i_{\tau_1} \dots i_{\tau_n} \\ l_1 \dots l_n}}
 \mathcal{M}_n^{\text{LO}}(\{t_i,i_{\tau_i}, l_i\}_{1 \leq i \leq n}).
\end{equation}
where the $ \mathcal{M}_n^{\text{LO}}(\{t_i,i_{\tau_i}, l_i\}_{1 \leq i \leq n})$ can once again be expressed as a product of determinants, their precise
form depending on the computed quantity.

This formalism leads to $4^n$ LO configurations for a given set of $n$ interaction times, to be compared
with $2^n$ possible configurations in the $\pm$ basis. However, we show in the next section that vacuum diagrams now directly
cancel in this formalism, without the actual need to perform an explicit sum over all configurations.

\subsection{Cancellation of vacuum diagrams}\label{subsec:cancellation}

In this section, we show the main result of this article: contributions to the partition function are
directly equal to zero in the LO basis. For all $n \geq 1$, $\{t_1, \dots, t_n\} \in [0,\tmax]^n$, $\{i_{\tau_1}, \dots, i_{\tau_n}\} \in \{-1,1\}^n$,
$\{l_1, \dots, l_n\} \in \{0,1\}^n$,
\begin{equation}\label{CancellationZ}
 Z_n^{\text{LO}}(\{t_i, i_{\tau_i},l_i\}_{1 \leq i \leq n}) =0,
\end{equation}
where the contributions to the partition function are
\begin{equation} 
  \begin{split}
    Z_n^{\text{LO}}(\{t_i, i_{\tau_i},l_i\}_{1 \leq i \leq n})& = \left(-\frac{i}{2}\right)^n i^n i^n \\
	& \hspace{-.8cm} \times \prod_\sigma \det \left[(\hat{\tilde g}_\sigma)_{L_i^\sigma L_j^\sigma}(t_i,t_j)\right]_{1 \leq i,j \leq n}.
\end{split}
\end{equation}
Each $-\frac{i}{2}$ comes from Eq. \eqref{lo_vertex} and the two $i^n$ factors from the fact that a straight line
actually represents an $i\hat{\tilde g}$ (Eq. \eqref{lo_g0}).

Let us consider an order $n \geq 1$ diagram contributing to $Z$. The interaction times are denoted $t_1, \dots, t_n$. We introduce $\hat{t}
= \mathrm{max}_i t_i$ and $\hat{i}$ such that $t_{\hat{i}} = \hat{t}$.
We label $\sigma$ the spin on the identity side of the $(\mathbf{1} \otimes \tau_\downarrow + \tau_\uparrow \otimes \mathbf{1})/2$ interaction 
vertex at $\hat{t}$, and $l$ the corresponding LO index.
In the consider, we consider the diagrammatic line following spin $\sigma$.
If $\hat{t}$ is surrounded by no other interaction vertex, the diagram is then proportional to 
\begin{equation}
(\hat{\tilde g}_\sigma)_{ll}(\hat{t},\hat{t}) = \delta_{l0} r_\sigma(\hat{t}, \hat{t}) + \delta_{l1} a_\sigma(\hat{t}, \hat{t}) = 0.
\end{equation}
In the case where $\hat{t}$ is surrounded by at least one other interaction vertex,
we label its surrounding interaction times (that can be equal) $t_i$ and $t_j$, $i,j \neq \hat{i}$,
with corresponding composite indices $L_i^\sigma$, $L_j^\sigma$.
We then obtain 
\begin{equation}
\begin{split}
(\hat{\tilde g}_\sigma)_{L_j^\sigma l}(t_j, \hat{t}) &= \delta_{L_j^\sigma 2} \delta_{l1} \left[\sigma k_\sigma(t_j, \hat t)
+ a_\sigma(t_j, \hat t)\right] \\
& + \delta_{L_j^\sigma 1}\delta_{l1}a_{\sigma}(t_j, \hat t) + \delta_{L_j^\sigma 0}\delta_{l1}k_\sigma(t_j, \hat t),
\end{split}
\end{equation}
and
\begin{equation}
  \begin{split}
(\hat{\tilde g}_\sigma)_{l L_i^\sigma}(\hat{t}, t_i) &= \delta_{L_i^\sigma 2} \delta_{l0} \left[r_\sigma(\hat t, t_i)
+ \sigma k_\sigma(\hat t, t_i)\right] \\
& + \delta_{L_i^\sigma 1}\delta_{l0}k_{\sigma}(\hat t, t_i) + \delta_{L_i^\sigma 0}\delta_{l0}r_\sigma(\hat t, t_i).
\end{split}
\end{equation}
The full diagram is then proportional to $\delta_{l0}\delta_{l1} = 0$.
Hence every diagram contributing to $Z$ in the LO basis is exactly equal to 0.
This formalism \emph{directly cancels vacuum diagrams}.

Finally, we note that this proof relies only on having the identity on one side of the interaction vertex, 
and not on the explicit contraction with $\tau_\uparrow, \tau_\downarrow$.
Had we kept the diagrammatics with $\hat g$ lines
instead of $\hat{\tilde g}$ ones, we would also obtain the cancellation of vacuum diagrams.
\subsection{Density computation and clusterization}\label{subsec:density}

In order to understand how to write the density of $\uparrow$ electrons on the energy level in the LO basis, we use the following property of the Keldysh formalism: the average value of an operator does
not depend on the branch of $\mathcal{C}$ where it is computed. Considering $d$ on the $+$ branch of the contour, the computation of the density
can be understood as the action of the 
$m_+$ matrix in the $\{ \pm\}$ basis, which transforms in the $\frac{1}{2} \tau_\uparrow$ matrix
in the LO basis according to Eq. \eqref{mp_eq}. Hence we can represent the measurement vertex as a ``special" interaction vertex at time $\tmax$ with $i_\tau = 1$:
\begin{fmffile}{lo_measurement}
\begin{equation}
\begin{gathered}
\begin{fmfgraph*}(50,20)
\fmfset{arrow_len}{3mm}
\fmfleft{i1}
\fmfright{o1}
\fmf{dashes_arrow,tension=2,label.dist=0, label=\vspace{.8cm} \hspace{.9cm} $\tau_\uparrow$}{i1,v1}
\fmf{dashes_arrow,tension=2,label.dist=0, label=\vspace{-.7cm} \hspace{-.9cm} $\tmax$}{v1,o1}
\fmffreeze
\fmfv{decor.shape=circle,decor.filled=shaded, decor.size=4thick}{v1}
\end{fmfgraph*}
\end{gathered}
\end{equation}
\end{fmffile}
\hspace{-.09cm}As previously, surrounding lines are dashed because they bear a $\hat{\tilde g}$ (and not an $i \hat{ \tilde g}$).
Hence the order-$n$ contribution to $d$ reads
\begin{equation} \label{d_lo}
\begin{split}
d_n = & \int_0^{\tmax} \mathrm{d}t_1 \dots \mathrm{d} t_n \sum_{\substack{i_{\tau_1} \dots i_{\tau_n} \\ l_1 \dots l_n}}\left(-\frac{i}{2}\right)^n \frac{i^{n+1}i^n}{i^2} \\
& \hspace{.5cm} \times \prod_\sigma \det \mathcal{D}_\sigma^{\text{LO}}(\{t_i, i_{\tau_i}, l_i\}_{1 \leq i \leq n}).
\end{split}
\end{equation}
The $\mathcal{D}_\sigma^{\text{LO}}$ matrices  are defined as
\begin{widetext}
  \begin{equation}\label{D_up_LO}
\mathcal{D}_\uparrow^{\text{LO}}(\{t_i,i_{\tau_i}, l_i\}_{1 \leq i \leq n}) = 
\begin{pmatrix}
 \left[(\hat{\tilde g}_\uparrow)_{L_i^\uparrow L_j^\uparrow}(t_i,t_j)\right]_{1 \leq i,j \leq n} & \begin{matrix}
(\hat{\tilde g}_\uparrow)_{L_1^\uparrow 2}(t_1, \tmax) \\
\vdots \\
(\hat{\tilde g}_\uparrow)_{L_n^\uparrow 2}(t_n, \tmax)
\end{matrix} \\
\begin{matrix}
(\hat{\tilde g}_\uparrow)_{2L_1^\uparrow}(\tmax, t_1) & \dots & (\hat{\tilde g}_\uparrow)_{2 L_n^\uparrow}(\tmax, t_n)
\end{matrix} & (\hat{\tilde g}_\uparrow)_{22}(\tmax, \tmax)
\end{pmatrix},
\end{equation}
\end{widetext}
and
\begin{equation}
\mathcal{D}_\downarrow^{\text{LO}}(\{t_i,i_{\tau_i}, l_i\}_{1 \leq i \leq n}) =\left[(\hat{\tilde g}_\downarrow)_{L_i^\downarrow L_j^\downarrow}(t_i,t_j)\right]_{1 \leq i,j \leq n}.
\end{equation}

Before considering the clusterization of interaction times, we note that half of the contributions to the density vanish.
Let us consider a given set $\{t_i, i_{\tau_i}, l_i\}_{1 \leq i \leq n}$ of LO vertices at order $n$, and let us label $\hat t = \max_i t_i$ and $\hat i$ such that $t_{\hat i} = \hat t$.
If $i_{\tau_{\hat i}} = 1$, then the $\downarrow$ spin is carrying the identity side of the vertex. As we measure the density on the $\uparrow$ spin, the argument used
in the cancellation vacuum diagrams (see \ref{subsec:cancellation}) applies again and $\mathcal{D}_\downarrow^{\text{LO}}(\{t_i,i_{\tau_i}, l_i\}_{1 \leq i \leq n})$ is the $n \times n$  null matrix.
If $i_{\tau_{\hat i}} = -1$, the contribution does not vanish.
Hence, when computing the density, at every order $n$ and for every set of $n$ interaction times, $4^n / 2$ LO configurations (out of $4^n$) are exactly zero.

The clusterization of interaction times around $\tmax$ in the calculation of the density is then a direct consequence of the cancellation of
vacuum diagrams and is very similar to the proof in the $\{\pm\}$ basis (now without the exponential sum).
Let $n$ be a given perturbation order, and $t_1 < t_2 < \dots < t_n$ $n$ interaction times. Let us assume that the first $j$ times are located far away from the measurement time
$t_\text{max}$, and that the last $n-j$ times are located in the vicinity of $t_\text{max}$. We can formally consider
\begin{equation}
\forall 1 \leq i \leq j, |t_i - t_\text{max}| \rightarrow  \infty.
\end{equation}
Because the Green's function is a local quantity in time, this means that for all $t \in \{t_1, \dots, t_j\}$, $t' \in \{t_{j+1}, \dots, t_n; \tmax\}$
\begin{equation}
  ||\hat{\tilde g}_\sigma(t , t')|| \rightarrow  0, \hspace{.2cm} ||\hat{\tilde g}_\sigma(t', t)|| \rightarrow  0.
\end{equation}

We therefore have
\begin{equation}
\prod_\sigma \det \mathcal{D}^{\text{LO}}_\sigma(\{t_i, i_{\tau_i}, l_i\}_{1 \leq i \leq n}) \simeq \prod_\sigma \det A_\sigma \prod_\sigma\det B_\sigma,
\end{equation}
with
\begin{subequations}
  \begin{gather}
  A_\sigma = \left[ (\hat{\tilde g})_{L_i^\sigma L_{i'}^\sigma}(t_i, t_{i'}) \right]_{1 \leq i, i'\leq j}, \\
  B_\downarrow = \left[ (\hat{\tilde g})_{L_i^\downarrow L_{i'}^\downarrow}(t_i, t_{i'}) \right]_{j+1 \leq i, i'\leq n} ,
\end{gather}
\end{subequations}
and $B_\uparrow$ is the $ \left[ (\hat{\tilde g})_{L_i^\uparrow L_{i'}^\uparrow}(t_i, t_{i'}) \right]_{j+1 \leq i, i'\leq n}$ matrix where a last line and column
corresponding to $\tmax$ are added, similar to Eq. \eqref{D_up_LO}.
However, $\prod_\sigma A_\sigma$ is a contribution to $Z$ at order $j$, and it vanishes according to (\ref{CancellationZ}). 
Therefore $\prod_\sigma \det D_\sigma \simeq 0$, and this proves the clusterization of times around $t_\text{max}$ in the computation of the density.

In the next section, we present different algorithms to stochastically sample Eqs~\eqref{d_keldysh} and~\eqref{d_lo}.

%
%
\section{Monte Carlo implementation} \label{sec:implementation}

In this section, we describe how to compute the density $d$ introduced above using quantum Monte Carlo (MC) techniques.
We present three different algorithms to compute this quantity, 
one using the $\pm$ algorithm presented in Ref.~\onlinecite{Profumo_prb_2015} and the other two based on the LO formalism
presented above.

\subsection{Monte Carlo algorithms}

We first describe how to stochastically generate MC configurations to sample the order-$n$ contribution, $d_n$, as expressed in Eqs \eqref{d_keldysh} and \eqref{d_lo}.

The $\pm$ algorithm works directly on the Keldysh contour. A configuration $\mathfrak{c}$ is determined by a given perturbation order $n$ and a set of $n$ interaction times (and \emph{not}
Keldysh points): $\mathfrak{c} = \{n ; t_1, \dots, t_n\}$. The contribution to $d_n$ of a given configuration
is
\begin{equation}
\begin{split}
w_\mathfrak{c}^\pm = -i^{n+1} &\sum_{\alpha_1 \dots \alpha_n}\alpha_1\dots\alpha_n \\
& \times \prod_\sigma \det \mathcal{D}_\sigma^\pm(\{t_i, \alpha_i\}_{1 \leq i \leq n}).
\end{split} 
\end{equation}
In the Monte Carlo, configurations are sampled stochastically according to their
weight, which we choose to be $| w_\mathfrak{c}^\pm |$. We then have
\begin{equation}
  d_n = \int_0^{\tmax} \mathrm{d}t_1 \dots \mathrm{d} t_n \, w_{\mathfrak{c}}^\pm \propto
  \sum^{\text{MC }\pm}_\mathfrak{c} \text{sign} \, w_{\mathfrak{c}}^\pm.
\end{equation}
Note that it was shown in Ref.~\onlinecite{Profumo_prb_2015} that $w_\mathfrak{c}^\pm \in \mathbb{R}$.

In the LO algorithm, a configuration $\mathfrak{c}$ is now determined by a given perturbation order $n$ and a set of $n$ interaction LO vertices:
 $\mathfrak{c} = \{n ; y_1, \dots, y_n\}$, where $y_i = \{t_i, i_{\tau_i}, l_i\}$.
Because the density is a real quantity, the contributions to $d_n$ of a configuration
$\mathfrak{c}$ can be written as
\begin{equation}
  w_\mathfrak{c}^{\text{LO}} = -\frac{1}{2^{n+1}}\text{Re} \left(i^{n+1} \prod_\sigma \det \mathcal{D}_\sigma^{\text{LO}}(\mathfrak{c})\right).
\end{equation}
If $|w_\mathfrak{c}^{\text{LO}}|$ is the statistical weight of $\mathfrak{c}$ in the Monte Carlo
process, then
\begin{equation}
  d_n = \int_0^{\tmax} \mathrm{d}t_1 \dots \mathrm{d} t_n \sum_{\substack{i_{\tau_1} \dots i_{\tau_n}\\ l_1 \dots l_n}} w_{\mathfrak{c}}^{\text{LO}}
   \propto \sum^{\text{MC LO}}_\mathfrak{c} \text{sign} \, w_{\mathfrak{c}}^{\text{LO}}.
\end{equation}

The third algorithm that we study is a mixed algorithm  that samples the configurations according to their
LO weight $|w_\mathfrak{c}^{\text{LO}}|$ but computes $d_n$ in the original $\{\pm\}$ basis, from the contributions $w_\mathfrak{c}^\pm$
at the sampled times.
A configuration $\mathfrak{c}$ is then determined by a given perturbation order $n$ and a set of $n$ interaction LO vertices:
 $\mathfrak{c} = \{n ; y_1, \dots, y_n\}$ and the MC weight is $|w_\mathfrak{c}^{\text{mixed}}| = |w_\mathfrak{c}^{\text{LO}}|$, so that
 \begin{equation}
\begin{split}
  d_n & = \frac{1}{\mathcal{N}} \int_0^{\tmax} \mathrm{d}t_1 \dots \mathrm{d} t_n \sum_{\substack{i_{\tau_1} \dots i_{\tau_n} \\
  l_1 \dots l_n}}\left|w_{\mathfrak{c}}^{\text{LO}}\right| \frac{ w_{\mathfrak{c}}^\pm}
{ \left|w_{\mathfrak{c}}^{\text{LO}}\right|} \\
& \propto \frac{1}{\mathcal{N}}\sum^{\text{MC mixed}}_\mathfrak{c} \frac{ w_{\mathfrak{c}}^\pm}{ \left|w_{\mathfrak{c}}^{\text{LO}}\right|},
\end{split}
\end{equation}
where $\mathcal{N}$ is the number of non-zero LO configurations. When computing the density, $\mathcal{N} = 4^n / 2$ at order $n$ (see Section \ref{subsec:density}).

In all three techniques, we use a standard Metropolis algorithm	\cite{metropolis_1953} to generate Markov chains distributed according to the weights $|w_{\mathfrak{c}}|$.
Starting from a given configuration $\mathfrak{c}$, a new configuration $\mathfrak{c}'$ is proposed according to one of the following two Monte Carlo updates:
\begin{enumerate}
\item Remove a randomly chosen interaction time (for the $\pm$ algorithm) or interaction LO vertex (for the LO and mixed algorithms) from $\mathfrak{c}$.
\item Add a new interaction time (for the $\pm$ algorithm) or an interaction LO vertex (for the LO and mixed algorithms). In all three techniques,
because of the clusterization of times around $\tmax$, we choose the new interaction time according to a Cauchy law (see below). We randomly
choose the $i_\tau$ and $l$ indices.
\end{enumerate}
The new configuration $\mathfrak{c}'$ is accepted or rejected with the usual Metropolis ratio
\begin{equation}
p_{\mathfrak{c} \rightarrow \mathfrak{c}'}^{\text{accept}} = \min\left(1, \frac{T_{\mathfrak{c}'\mathfrak{c}}|w_{\mathfrak{c}'}|}{T_{\mathfrak{c}\mathfrak{c}'}|w_{\mathfrak{c}}|}\right),
\end{equation}
where $T_{\mathfrak{c}\mathfrak{c}'}$ is the probability to propose $\mathfrak{c}'$ after $\mathfrak{c}$.

\subsection{Proposition of times}

We have shown previously that times clusterize around $\tmax$. It is therefore more efficient to propose times located around it compared to uniformly
distributed between $0$ and $\tmax$. We consider a Cauchy law determined by two parameters $t_0$ and $a$
\begin{equation}
\rho(t)= \frac{1}{C} \frac{1}{1 + \left(\frac{t-t_0}{a}\right)^2}.
\end{equation}
$C$ is a normalization factor such that the integral of $\rho$ between $0$ and $\tmax$ gives 1, defined as
$C = a\left[ C_2 - C_1\right]$,
where $C_1 =  \arctan\left(-\frac{t_0}{a}\right)$ and $C_2 = \arctan \left(\frac{\tmax - t_0}{a}\right)$.

To obtain a new time that follows this probability law, one can perform these three steps: 
\begin{enumerate}
\item Choose a random number $u$ uniformly distributed between 0 and 1.
\item Construct
\begin{equation}
x = \frac{1}{2} + \frac{1}{\pi}\left[ (1-u)C_1 + u C_2\right],
\end{equation}
 uniformly distributed between $\frac{1}{2} + \frac{1}{\pi}C_1$ and $\frac{1}{2} + \frac{1}{\pi}C_2$.
\item Compute
\begin{equation}
t = t_0 + a \tan\left(\pi \left(x - \frac{1}{2}\right)\right),
\end{equation}
distributed between $0$ and $\tmax$ according to $\rho$.
\end{enumerate}

The parameters $t_0$ and $a$ are then fitted to the 1D projection of times visited by the Monte Carlo,
accumulated during the first part of the computation.

\subsection{Redefinition of noninteracting propagators}\label{subsec:alpha_shift}
 
As shown in previous works\cite{rubtsov_prb_2005, wei_prb_2017, Profumo_prb_2015,
rossi_prb_2016}, there is some freedom in the choice of the
noninteracting propagator used to construct the perturbation expansion,
since the interaction can be redefined as
\begin{equation}\label{definitionAlpha}
  U n_\uparrow n_\downarrow = U(n_\uparrow - \alpha)(n_\downarrow - \alpha)
  + U \alpha (n_\uparrow + n_\downarrow) + \text{const}.
\end{equation}
Note that in this subsection $\alpha$ \emph{does not} denote a Keldysh
index but a scalar, in order to be consistent with the existing literature.
In particular, it was shown that $\alpha$ can strongly modify the radius of convergence of the perturbation series \cite{Profumo_prb_2015, wei_prb_2017}.
This redefinition of the interaction term in Eq. \eqref{definitionAlpha} 
is taken into account by subtracting $\alpha$ on the diagonal of the determinants 
as explained and proved in Ref. \onlinecite{Profumo_prb_2015}.
The second term in Eq. \eqref{definitionAlpha} acts as a shift in the chemical potential and can be absorbed in a redefinition of the noninteracting propagators.

Let us first consider the LO basis. This shift acts a diagonal term in the self-energy and hence in
\begin{equation}
  \left(\hat g_\sigma^{\text{LO}}\right)^{-1} = \begin{pmatrix}
  r_\sigma^{-1} & -k_\sigma / |r_\sigma|^2 \\
  0 & a_\sigma^{-1}
\end{pmatrix}.
\end{equation}
$\alpha$ therefore modifies $r_\sigma$ and $a_\sigma$ into  
\begin{align}\label{redef_r}
  \bar r_\sigma(\omega) &= \left[ r(\omega)^{-1}- U\alpha \right]^{-1}, \\
  \bar a_\sigma (\omega) &= \left[ a(\omega)^{-1}- U\alpha \right]^{-1}.
\end{align}
As $k_\sigma / |r_\sigma|^2$ is not impacted by the shift, the
modified Keldysh Green's function is then
\begin{equation}
\bar k_\sigma(\omega)= \left| \frac{\bar r_\sigma(\omega)}{r_\sigma(\omega)}\right|^2 k_\sigma(\omega).
\end{equation}
From these expressions, we can then deduce the modified Green's functions in the $\{\pm\}$ basis through a change of basis tranformation.

\subsection{Normalization procedure}\label{subsec:normalisation}

All Monte Carlo algorithms presented above compute the order-$n$ contribution to the density $d$, however the MC results need to be normalized.
Hence we restrict our calculation to two consecutive orders, $n$ and $n + 1$,  and a time or vertex can be added (resp. removed) only if the current
configuration $\mathfrak{c}$
is at order $n$ (resp $n+1$). We measure both the density ($d_{n}$ and $d_{n+1}$) and a 
normalization factor ($\eta_n$ and $\eta_{n+1}$). 
In all algorithms, 
the normalization factor is chosen to be the sum of the absolute value of the contributions
to the density:
\begin{equation}
\eta_n \propto \sum^{\text{MC}}_\mathfrak{c}  | w_{\mathfrak{c}} |,
\end{equation}
where the proportionality constant is the same as in the calculation of
$d_n$. If $\tilde{d}_n$ and $\tilde{\eta}_n$ are the unrenormalized sums of the
contributions accumulated in the Monte Carlo procedure, then the
normalized values for $d_n$ and $\eta_n$ are obtained as
\begin{equation}\label{eq:normalization}
d_{n+1} = \frac{\eta_{n}}{\tilde\eta_{n}}  \tilde{d}_{n+1}; \text{\hspace{.5cm}}
\eta_{n+1} = \frac{\eta_{n}}{\tilde\eta_{n}} \tilde\eta_{n+1},
\end{equation}
and $\eta_n$ is then used to normalize the following simulation between orders $n + 1$ and $n+2$. 
The lowest order is computed analytically to close the equations.

%
%
\section{Results} \label{sec:results}

\subsection{Density}
\begin{figure}
\centering
\includegraphics[width=0.45\textwidth]{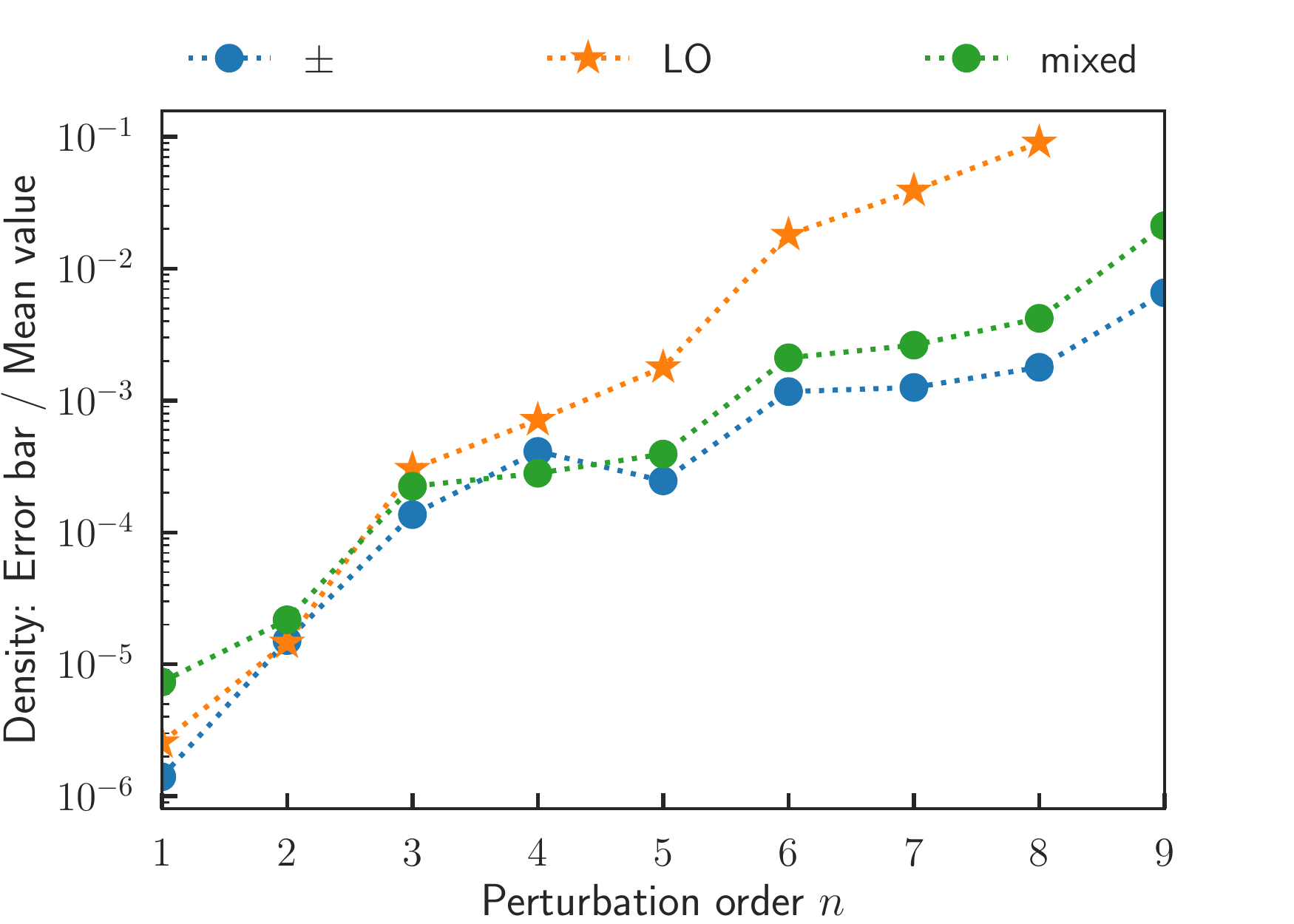}
  \caption{Comparison of the error bar divided by the mean value in a density computation, for the three different MC algorithms considered: the one working in the Keldysh $\pm$ basis (blue dots),
  the one in the LO basis (orange stars) and the mixed algorithm (green dots, see text). $t=1$, $\beta t = 100$, $\gamma^2 = 0.04 t^2$, $\epsilon_d = -0.36 t$, $U=1.2 t$,
  $\alpha = 0.3$. Computational effort is 240 CPU*hours for every order.
}
\label{EB_Mean}
\end{figure}

In this section, we present actual computations of the density according to the algorithms described in the previous section and compare their
efficiency. In the following, we consider
an energy level $\epsilon_d$ coupled to a bath described by a semi-circular density of states
of bandwidth $4t$. The Green's function describing this bath is defined
on the complex plane as \cite{RevModPhys.68.13}
\begin{equation}
g_{\mathrm{bath}}(\zeta)  = \frac{\zeta - \mathrm{sgn}(\mathrm{Im}\zeta)\sqrt{\zeta^2-4t^2}}{2t^2}.
\end{equation}
The noninteracting retarded Green's function of the impurity level is
\begin{equation}
r_\sigma(\omega) = \frac{1}{\omega - \epsilon_d - \gamma^2  g_{\mathrm{bath}}(\omega)},
\end{equation}
where $\gamma$ is a coupling term between the energy level and the bath. The Keldysh Green's function is then deduced using the fluctuation-dissipation theorem
\begin{equation}
  k_\sigma(\omega) = \tanh\left(\frac{\beta \omega}{2}\right) \left[r_\sigma(\omega) - r^*_\sigma(\omega)\right].
\end{equation}

In the following, $t = 1$ is our energy unit. We consider $\beta t = 100$, $\gamma^2 = 0.04 t^2$, $\epsilon_d = -0.36 t$. Electrons on the impurity experience a 
local Coulomb interaction $U = 1.2 t$. We choose the $\alpha$ shift to be $\alpha = 0.3$ (see Sec.~\ref{subsec:alpha_shift}), such that $U \alpha = -\epsilon_d$.
The bath being particle-hole symmetric, this creates a shifted retarded Green's function $\bar r(\omega)$ that is itself particle-hole symmetric (see Eq.~\eqref{redef_r}).
However, we have checked that this particular choice of $\alpha$ does not influence our conclusions.
We provide in Appendix~\ref{app:benchmark} a table benchmarking the LO and mixed algorithms against the original $\pm$ algorithm.
This shows in particular that the LO and mixed algorithms yield correct results
and that we can indeed reach long times in the LO algorithm without an exponential sum of determinants.

Our main result is shown on Figure~\ref{EB_Mean} where we compare the relative error bar in 
the density computation as a function of the perturbation order. Blue dots denote the $\pm$
algorithm, orange stars the LO algorithm, and green dots the mixed algorithm. The order-9 relative error is not shown for the LO algorithm as it exceeds 1
and is therefore meaningless.
In all three cases, dotted lines are guides to the eye. The computational time is 240 CPU*hours for each order.

We see that all three relative error bars increase with perturbation order.
This can either come from the increasing difficulty of computing the series
coefficients, or an error propagation coming from the normalization factor
$\eta$. We plot in Appendix \ref{app:eb} the relative error bar on $\eta$,
which is much smaller than the final relative error on the density, showing
that the latter mainly comes from the increasing difficulty to compute higher order
coefficients. Moreover, the LO relative error bars very quickly
become much larger than the $\pm$ ones, their difference nearly reaching two
orders of magnitude at order $8$.  The mixed algorithm is found to perform
better than the LO algorithm but its error bars slowly grow larger than the
$\pm$ ones. This is surprising, as one could have expected to at least gain the
decorrelation time over the algorithm of Ref.~\onlinecite{Profumo_prb_2015}.
We discuss the origin of the error bars in  both algorithms in the next section.


\subsection{The return of the sign problem}

\begin{figure}
\centering
\includegraphics[width=0.49\textwidth]{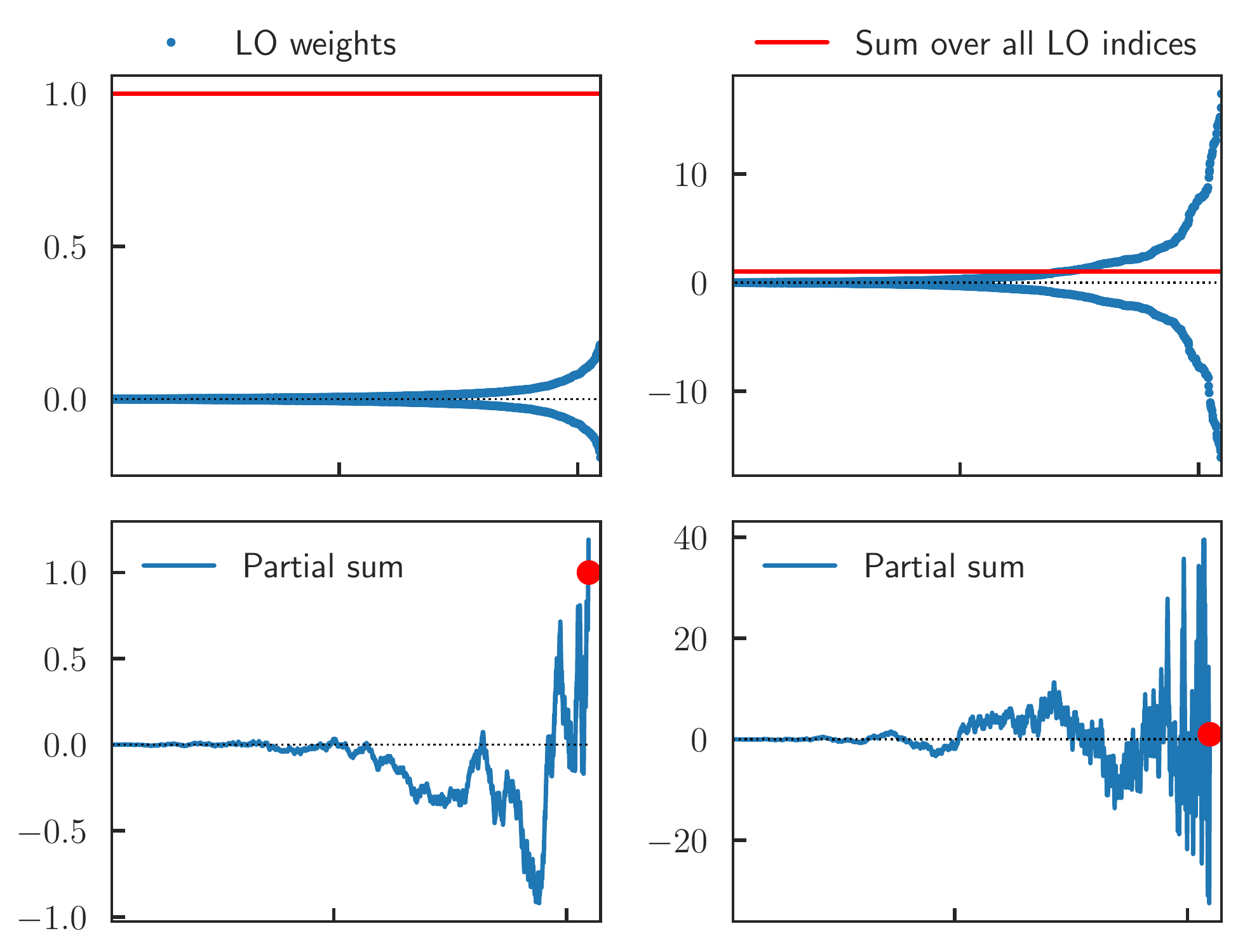}
  \caption{\emph{Upper panel}: Sorted array of the LO weights according to their absolute value (blue dots)  and their sum (red line), normalized to 1.
\emph{Lower panel}: Partial sum of the above LO weights, from left to right, the red dot being the last point, by definition 1.
Left panels correspond to the $T_1 = \{ 273.2, 277.8, 280.9, 331.7, 366.4, 390.5\}$ time configuration, and the right panels to
$T_2 = \{ 338.3, 343.2, 366.9, 369.7, 393.9, 394.5\}$. Order 7, $\tmax = 400$.
}
\label{LO_weights}
\end{figure}

In this section, we discuss the origin of the large variance in the computation of the density in the LO algorithm
in terms of a sign problem in the Monte Carlo sampling and we show how this impacts the error bars of
the mixed algorithm.

On the upper panel of Figure \ref{LO_weights}, we plot as blue dots the non-zero LO weights for two different time configurations, sorted according to their
absolute value.
The left and right panel correspond to two different time configurations (Cf caption).
In both cases, the red line indicates the full sum over all LO
indices, normalized to 1 (which coincides with the $\pm$ weights). The lower panel shows the partial sum, from left to right, of the LO weights plotted above.
The last point, equal to 1 by construction, is emphasized as a red dot.
As roughly half of the weights are positive and half negative, we see that the sum of the 
LO weights over the indices at fixed time configuration is characterized by a massive cancellation.
This is the origin of the large error bar in the Monte-Carlo, \emph{i.e.} another manifestation of the sign problem.
Furthermore, the partial sum shows that there is no clear feature or cutoff from which one could extract the value of the full sum.

\begin{figure}
\centering
\includegraphics[width=0.49\textwidth]{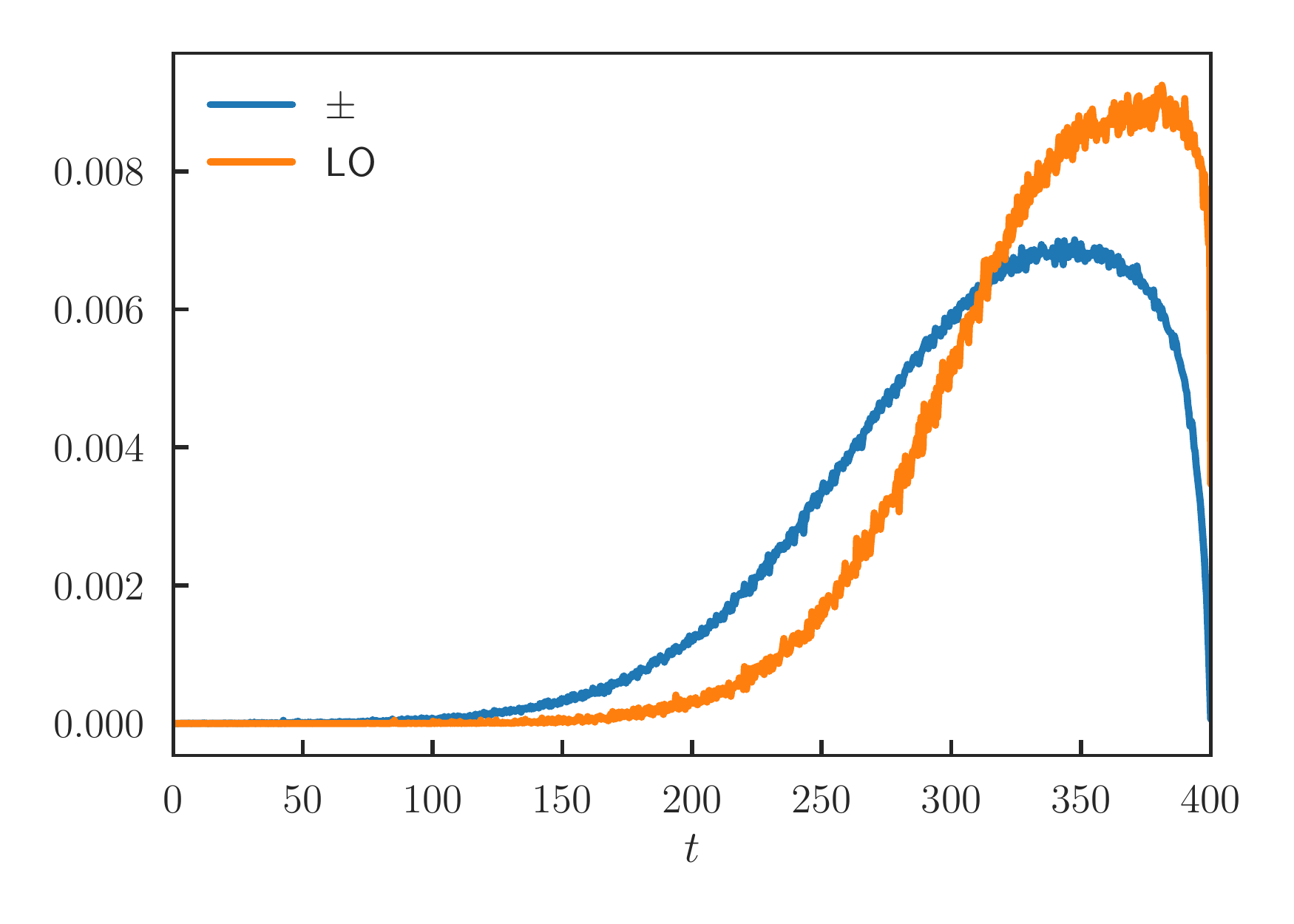}
\caption{Histograms of the times visited by the Monte Carlo algorithms, projected in one dimension. Order 9, $\tmax = 400$.}
\label{Histo_times}
\end{figure}

Let us now turn to the mixed algorithm.
On both the left and right panels of Figure \ref{LO_weights}, the sum over all LO indices, which coincides with the $\pm$ weight, is normalized to 1.
However, on the left panel, the weights of the different LO configurations are small compared to the final result, reaching at most 20\% of it.
On the right panel, those same weights are much bigger, reaching up to 1700\% of the full sum. Hence the Monte Carlo implemented
in the LO basis does not sample the same time configurations as the algorithm in the $\{\pm\}$ basis.
This is illustrated in Figure \ref{Histo_times} where the histograms of the times visited by the Monte Carlo, projected in one dimension, are plotted for both the $\pm$ algorithm (blue line)
and LO one (orange line).
First, we observe the clusterization of times proved at the beginning of this article: interaction times contributing to the density tend to be in the vicinity of $\tmax$.
Then, we see that some times located far away from the measurement but still contributing significantly to the $\pm$ algorithm
are almost never visited in the LO algorithm.
On the other hand, times close to $t_\text{max}$ are more sampled in the latter.
As  times visited by the mixed algorithm coincide with the LO ones, 
this explains the difference in error bars between the mixed and $\pm$ algorithms observed in Figure \ref{EB_Mean}.

%
%

\section{Conclusion}\label{sec:conclusion}

In conclusion, the explicit sum over the Keldysh indices of the original $\pm$ algorithm of Ref.~\onlinecite{Profumo_prb_2015}
has two functions: {\it i)} it allows to reach the very long times due to the clusterization of the integrand caused
by the cancellation of vacuum diagrams;
{\it ii) } it strongly reduces the error bar by performing a massive cancellation of terms.
In this article, we have shown that one can obtain the first properties for each determinant 
using the Larkin-Ovchinnikov basis, hence   without the exponentially large sum of determinants.
A direct implementation of the algorithm in the LO basis
indeed reaches the steady state, but also has an error bar growing quickly with the order $n$ due to a sign problem.
An interesting possibility would be the existence of an optimum between the LO and original $\pm$ algorithms, 
using partial groupings of terms in the LO basis with less than $2^n$ terms that would reduce the sign problem
and yields a better scaling than the original algorithm in the $\{\pm\}$ basis.
Work is in progress in this direction.



\begin{acknowledgments}
We are grateful to Xavier Waintal and Antoine Georges for useful discussions.
This work was partly supported by the European Research Council grant ERC-278472-MottMetals (PS, OP). 
The Flatiron Institute is a division of the Simons Foundation.
Part of this work was performed using HPC resources from GENCI (Grant No. A0050510609).
\end{acknowledgments}


\begin{appendix}

  \section{Clusterization of the density in the $\{\pm\}$ basis}\label{app:clusterization}

  We reproduce here the argument of Ref. \onlinecite{Profumo_prb_2015} showing that the cancellation of vacuum diagrams when summing
  over Keldysh indices implies the clusterization of interaction times near $\tmax$.

Let $n$ be a given perturbation order, and $t_1 < t_2 < \dots < t_n$ $n$ interaction times. Let's assume that the first $j$ times are located far away from the measurement time
$t_\text{max}$, and that the last $n-j$ times are located in the vicinity of $t_\text{max}$. We can formally consider
\begin{equation}
\forall 1 \leq i \leq j, |t_i - t_\text{max}| \rightarrow  \infty.
\end{equation}
Because the Green's function is a local quantity, this means that for all $t \in \{t_1, \dots, t_j\}$, $t' \in \{t_{j+1}, \dots, t_n; \tmax\}$
\begin{equation}
  ||\hat g_\sigma(t , t')|| \rightarrow  0, \hspace{.2cm} ||\hat g_\sigma(t', t)|| \rightarrow  0.
\end{equation}

We therefore have
\begin{equation}
  \begin{split}
    & \sum_{\alpha_1 \dots \alpha_n}  \alpha_1 \dots \alpha_n  \prod_\sigma \det \mathcal{D}^\pm_\sigma(\{t_i, i_{\tau_i}, l_i\}_{1 \leq i \leq n}) \\
    \simeq & \sum_{\alpha_1 \dots \alpha_j}  \alpha_1 \dots \alpha_j \prod_\sigma \det A_\sigma \\
   & \hspace{.5cm}\times \sum_{\alpha_{j+1} \dots \alpha_n} \alpha_{j+1} \dots \alpha_n  \prod_\sigma \det B_\sigma,
  \end{split}
\end{equation}
with
\begin{subequations}
  \begin{gather}
    A_\sigma = \left[ (\hat g_\sigma)_{\alpha_i \alpha'_i}(t_i, t_{i'}) \right]_{1 \leq i, i'\leq j}, \\
    B_\downarrow = \left[ (\hat g_\downarrow)_{\alpha_i \alpha'_i}(t_i, t_{i'}) \right]_{j+1 \leq i, i'\leq n} ,
\end{gather}
\end{subequations}
  and $B_\uparrow$ is the $ \left[ (\hat g_\uparrow)_{\alpha_i \alpha'_i}(t_i, t_{i'}) \right]_{j+1 \leq i, i'\leq n}$ matrix where a last line and column
corresponding to $\tmax$ are added, similar to Eq. \eqref{D_up_pm}.
However, $ \sum_{\alpha_1 \dots \alpha_j} \alpha_1 \dots \alpha_j  \prod_\sigma A_\sigma$ is a contribution to $Z$ at order $j$, and it vanishes according to (\ref{CancellationZ_pm}). 
Therefore 
 $\sum_{\alpha_1 \dots \alpha_n} \alpha_1 \dots \alpha_n \prod_\sigma \det D^\pm_\sigma \simeq 0$,
 and this proves the clusterization of times in around $t_\text{max}$ in the computation of the density.

\section{Benchmark}\label{app:benchmark}

The table below benchmarks the contributions to the density between the $\pm$, LO, and mixed algorithms.
  We take $t=1$ as our energy unit, and parameters are $\beta t=100$, $\gamma^2 = 0.04 t^2$, $\epsilon_d = -0.36 t$, $U=1.2 t$, $\alpha=0.3$.
  Computation effort is 240 CPU*hours for each perturbation order.
  \begin{widetext}
\begin{center}
  \begin{tabular}{ | l | c | c | c |}
    \hline
    &&&\\[-1em]
    & $\pm$ & LO & mixed \\ \hline
    &&&\\[-1em]
    Order 1 & $ -1.7013454 \pm 0.00014\%$ & $-1.7013431 \pm 0.00026\%$ & $-1.7013466 \pm 0.00073\%$ \\ \hline
    &&&\\[-1em]
    Order 2 & $14.47243 \pm 0.0015\%$ & $14.47252 \pm 0.0015\%$ & $14.47214 \pm 0.0022\%$ \\ \hline
    &&&\\[-1em]
    Order 3 & $ -33.3479 \pm 0.014\%$ & $-33.3610 \pm 0.030\%$ & $-33.3583 \pm 0.022\%$\\ \hline
    &&&\\[-1em]
    Order 4 & $-431.09 \pm 0.041\%$ & $-431.51 \pm 0.071\%$ & $-431.30 \pm 0.028\%$\\ \hline
    &&&\\[-1em]
    Order 5 & $5094.7 \pm 0.025\%$ & $5100.6 \pm 0.18\%$ & $5092.6 \pm 0.039\%$\\ \hline
    &&&\\[-1em]
    Order 6 & $-16173 \pm 0.12\%$ & $-15802 \pm 1.8\%$ & $-16171 \pm 0.21\%$\\ \hline
    &&&\\[-1em]
    Order 7 & $-1.6411 \times 10^{5} \pm 0.13\%$ & $-1.6595 \times 10^5 \pm 3.9\%$ & $-1.6554 \times 10^5 \pm 0.26\%$\\ \hline
    &&&\\[-1em]
    Order 8 & $2.2332 \times 10^7 \pm 0.18 \%$ & $2.1071 \times 10^7 \pm 9.0\%$ & $2.2316 \times 10^7 \pm 0.42\%$\\ \hline
    &&&\\[-1em]
    Order 9 & $-7.865 \times 10^7 \pm 0.66\%$ & $2.852 \times 10^7 \pm 240\%$ & $-8.079 \times 10^7 \pm 2.1\%$\\
    \hline
  \end{tabular}
\end{center}
  \end{widetext}


\section{Origin of error bar}\label{app:eb}

\begin{figure}
\centering
\includegraphics[width=0.45\textwidth]{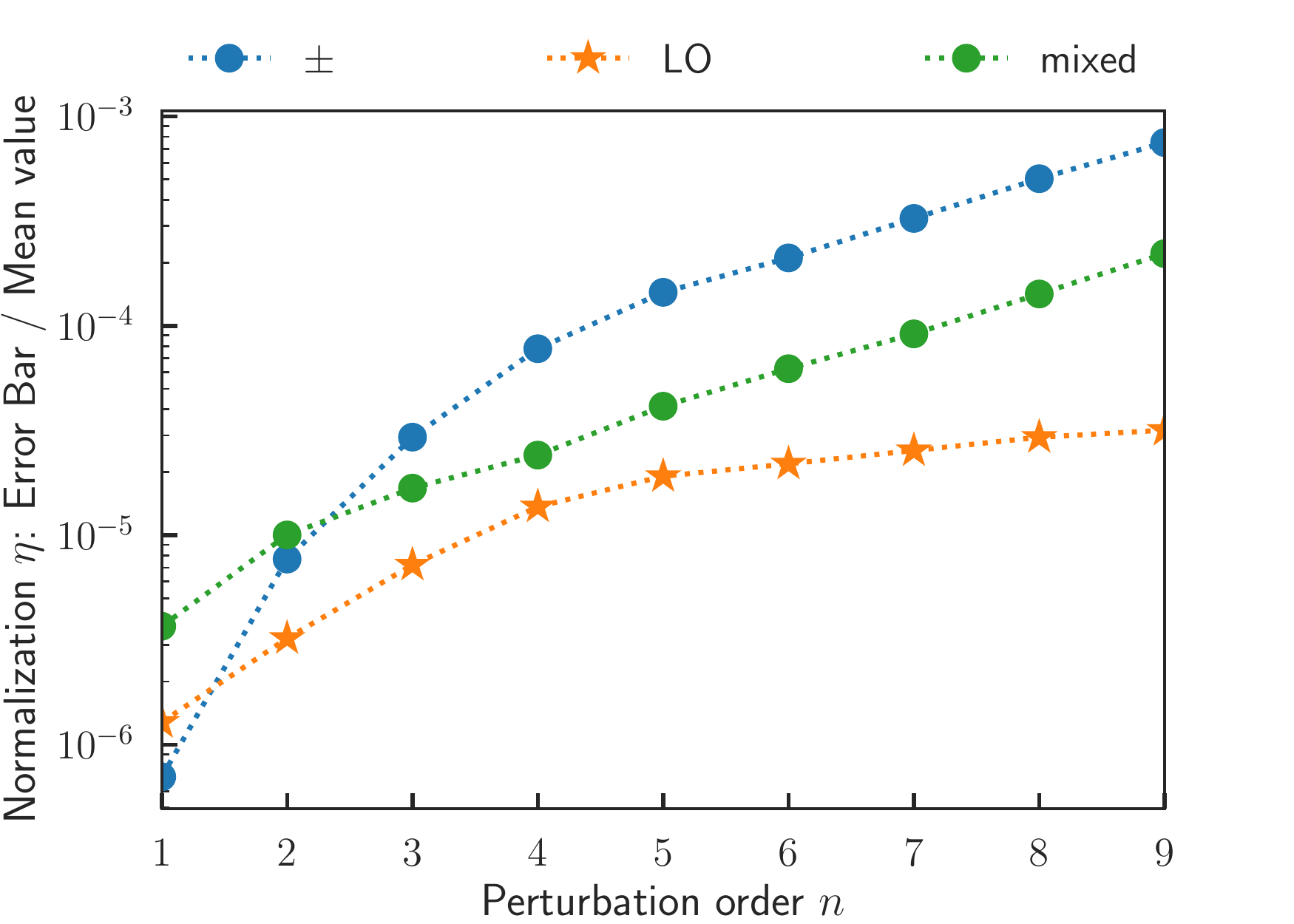}
\caption{Comparison of the error bar divided by the mean value of the normalization coefficient,
 for the three different MC algorithms considered: the one working in the Keldysh $\pm$ basis (blue dots),
  the one in the LO basis (orange stars) and the mixed algorithm (green dots). $t=1$, $\beta t = 100$, $\gamma^2 = 0.04 t^2$, $\epsilon_d = -0.36 t$, $U=1.2 t$,
  $\alpha = 0.3$. Computational effort is 240 CPU*hours for every order.
}
\label{Norm_coeff_EB}
\end{figure}

We have seen in Sec. \ref{subsec:normalisation} that the contributions to the density have to be normalized by a factor $\eta$, see Eq. \eqref{eq:normalization}.
To verify that the error bars on the density are not due to this normalization factor, we plot its relative error bars on Figure \ref{Norm_coeff_EB}.
Blue dots denote the $\pm$ algorithm, orange stars the LO algorithm, and green dots the mixed algorithm. Comparing it to Figure \ref{EB_Mean}, we see that the relative
error bars on $\eta$ are much smaller than the ones on the density.

\end{appendix}

\bibliography{main}

\end{document}